\documentclass[showpacs,amsmath,amssymb,aps,pra,
longbibliography,
10pt,reprint,superscriptaddress]{revtex4-1}
\usepackage{bm}
\usepackage[breaklinks=true,colorlinks=true,linkcolor=blue,urlcolor=blue,citecolor=blue]{hyperref}
\usepackage{dcolumn}
\usepackage{amsmath,amssymb}
\usepackage{mathdots}
\usepackage{natbib}
\usepackage{soul,color}
\usepackage{graphicx}
\usepackage{epstopdf}
\usepackage[note-name]{notes2bib}
\usepackage{mathptmx}

\begin{document}

\title{Vortex counting and velocimetry for slitted superconducting thin strips}

\author{V.\,M. Bevz}
    \affiliation{Physics Department, V. Karazin Kharkiv National University, 61022, Kharkiv, Ukraine}
\author{M.\,Yu.~Mikhailov}
    \affiliation{B. Verkin Institute for Low Temperature Physics and Engineering of\\
     the National Academy of Sciences of Ukraine, 61103, Kharkiv, Ukraine}
\author{B.~Budinsk\'a}
    \affiliation{University of Vienna, Faculty of Physics, 1090 Vienna, Austria}
    \affiliation{University of Vienna, Vienna Doctoral School in Physics, 1090 Vienna, Austria}
\author{S.~Lamb-Camarena}
    \affiliation{University of Vienna, Faculty of Physics, 1090 Vienna, Austria}
    \affiliation{University of Vienna, Vienna Doctoral School in Physics, 1090 Vienna, Austria}
\author{S.\,O.~Shpilinska}
    \affiliation{University of Vienna, Faculty of Computer Science, 1090 Vienna, Austria}
\author{A.\,V.~Chumak}
    \affiliation{University of Vienna, Faculty of Physics, 1090 Vienna, Austria}
\author{M.~Urb\'anek}
    \affiliation{CEITEC, Brno University of Technology, 61200 Brno, Czech Republic}
\author{M. Arndt}
    \affiliation{University of Vienna, Faculty of Physics, 1090 Vienna, Austria}
\author{W.\,Lang}
    \affiliation{University of Vienna, Faculty of Physics, 1090 Vienna, Austria}
\author{O.\,V. Dobrovolskiy}
    \email[Corresponding author:]{oleksandr.dobrovolskiy@univie.ac.at}
    \affiliation{University of Vienna, Faculty of Physics, 1090 Vienna, Austria}
\date{\today}

\begin{abstract}
The maximal speed $v^\ast$ for magnetic flux quanta is determined by the energy relaxation of unpaired electrons and is thus essential for superconducting microstrip single-photon detectors (SMSPDs). However, the deduction of $v^\ast$ from the current-voltage ($I$-$V$) curves at zero magnetic field is hindered by the unknown number of vortices, $n_\mathrm{v}$, as a small number of fast-moving vortices can induce the same voltage as a large number of slow-moving ones. Here, we introduce an approach for the quantitative determination of $n_\mathrm{v}$ and $v^\ast$. The idea is based on the Aslamazov and Larkin prediction of kinks in the $I$-$V$ curves of wide and short superconducting constrictions when the number of fluxons crossing the constriction is increased by one. We realize such conditions in wide MoSi thin strips with slits milled by a focused ion beam and reveal quantum effects in a macroscopic system. By observing kinks in the $I$-$V$ curves with increase of the transport current, we evidence a crossover from a single- to multi-fluxon dynamics and deduce $v^\ast\simeq12\,$km/s. Our experimental observations are augmented with numerical modeling results which reveal a transition from a vortex chain over a vortex jet to a vortex river with increase of $n_\mathrm{v}$ and the vortex velocity. Our findings are essential for the development of 1D and 2D few-fluxon devices and provide a demanded approach for the deduction of $v^\ast$ at the SMSPD operation conditions.
\end{abstract}
\maketitle

\maketitle

\section{Introduction}
The ultimate speed limit for magnetic flux transport in superconductors via quantized flux lines (fluxons) is attracting increasing attention\,\cite{Wor12prb,Gri15prb,Jel16nsr,Emb17nac,She17prb,Shk17prb,Kog20prb,Dob20pra,Pat21prb,Kog22prb,Dob20nac,Bud22pra,Dob23inb}. This interest is caused by the fundamental questions regarding the interaction and stability of vortices as topological objects moving at $>10$\,km/s that exceed the maximal drift velocity of the Cooper-pair condensate\,\cite{Emb17nac,Kog20prb}. On the applied side, ultrafast vortex dynamics underlies the vortex-assisted voltage response of superconducting microstrip single-photon detectors (SMSPDs)\,\cite{Kor20pra,Cha20apl,Chi20apl}. When operated at a close-to-depairing current $I \lesssim I_\mathrm{dep}$, the intrinsic detection efficiency of SMSPDs is predicted to reach almost unity\,\cite{Vod17pra}, opening novel prospects for large-active-area detectors in free-space quantum cryptography\,\cite{Che21nat}, macromolecule analysis\,\cite{Fei19nph}, and other areas\,\cite{Ste21apl}.

The maximal vortex velocities $v^\ast$ deduced from current-voltage ($I$-$V$) curves are used for judging whether a material could be potentially suitable for SMSPDs\,\cite{Lin13prb,Cap17apl,Hof21tsf,Liu21sst,Cir21prm}. This
approach is based on the theory of flux-flow instability (FFI)\,\cite{Lar75etp,Bez92pcs}, in which $v^\ast$ depends on the energy relaxation time of unpaired electrons (quasiparticles) $\tau_\varepsilon$\,\cite{Per05prb,Leo11prb,Att12pcm,Dob19pra,Leo20sst}. Thus, FFI studies represent an alternative method for material characterization, which is complementary to photoresponse, magnetoconductance, and microwave measurements\,\cite{Wor12prb,Dob15apl,Pom08prb,Loe19acs,Pok22jap,Zha16prb,Sid21prb}. The temperature dependence $\tau_\varepsilon(T)$ contains information on the dominating microscopic mechanism of
the quasiparticle relaxation\,\cite{Kap76prb,Per05prb,Arm07prb,Leo11prb,Att12pcm,Cap17prb}, including electron-electron\,\cite{Doe94prl} and electron-phonon\,\cite{Per05prb,Emb17nac} scattering, recombination\,\cite{Per05prb,Leo11prb}, and escape of nonequilibrium phonons to the substrate\,\cite{Vod19sst,Dob20nac}. An accurate determination of $v^\ast$ and $\tau_\varepsilon$ is thus essential for basic research and the development of applications.

However, the deduction of $v^\ast$ at low applied magnetic field $B$ is difficullt since the local magnetic field in a current-carrying conductor can differ from $B$. Thus, for most SMSPD-suitable materials, such as WSi\,\cite{Zha16prb,Chi20apl}, MoSi\,\cite{Kor20pra,Cha20apl,Bud22pra}, MoN\,\cite{Ust20jpd,Ust20sst}, NbC\,\cite{Dob20nac}, NbRe\,\cite{Ejr22apl}, NbReN\,\cite{Cir21prm}, and NbTiN\,\cite{Sid21prb}, the dependence $v^\ast(B)$ will likely diverge at $B\rightarrow0$ if one uses the standard relation\,\cite{Bra95rpp}
\begin{equation}
\label{e1}
    v^\ast = \displaystyle\frac{V^\ast}{BL},
\end{equation}
where $V^\ast$ is the voltage measured just before the jump to the highly resistive state and $L$ is the distance between the voltage leads. Given that $\tau_\varepsilon$ is finite, this divergence is obviously unphysical and it is caused by the failure of Eq.\,\eqref{e1} at $B\rightarrow0$, as the number of vortices $n_\mathrm{v}$ in the superconductor can be \emph{larger} than follows from the estimate $n\Phi_0 = BS$\,\cite{Dob20nac} [$n=1,2,..$; $\Phi_0$: magnetic flux quantum; $S$: area of the sample surface]. The larger number of vortices is caused by the current flowing through the strip, which leads to the nucleation of vortices and, thus, the resistive state associated with the current-driven vortex motion\,\cite{Bra95rpp}. While the current-induced generation of vortices is insignificant at large magnetic fields, this mechanism becomes dominant at low fields, turning into the only source of vortices (and antivortices) at $B=0$. In this way, the highest vortex velocities are expected in the low/zero-$B$ regime and the key challenge consists in determining the number of fluxons moving in the superconductor.

For straight strips, the problem of vortex counting and velocimetry is complicated due to simultaneous vortex entry from many edge defects\,\cite{Bud22pra}. In contrast, an artificially introduced single edge defect can be used for suppressing the energy barrier locally and letting the vortices enter at a deliberately chosen edge point\,\cite{Ala01pcs,Bez22prb}. As the current increases, this point turns into an FFI nucleus\,\cite{Vod19sst}, with a vortex river (chain of vortices with a reduced number of quasiparticles in their cores) spreading across the superconductor and causing a local FFI\,\cite{Bez19prb,Bud22pra}. Herein, the average number of vortices in the strip $n_\mathrm{v}$ is determined by the balance between the intervortex repulsion and the current-vortex and vortex-edge interactions\,\cite{Bez22prb,Asl75etp}. However, if one uses the Josephson relation $V^\ast = n_\mathrm{v} \Phi_0 f^\ast$, where $f^\ast = v^\ast/w$ is the frequency of the vortex entry into the superconductor of width $w$, the unknown $n_\mathrm{v}$ still restricts the analysis to finding the  proportionality $V^\ast\propto f^\ast\propto v^\ast$\,\cite{Emb17nac,Bez22prb} rather than the absolute value of $v^\ast$. This restriction is associated with the uncertainty of whether the voltage generated by moving vortices arises due to a \emph{smaller number} of \emph{fast-moving} vortices or due to a \emph{larger number} of \emph{slow-moving} ones. Hence, a technique for controlling the number of vortices is required for the deduction of $v^\ast$ and for the development of superconducting devices operating in the few- and single-fluxon regime.

Here, we introduce an approach for the quantitative determination of $n_\mathrm{v}$ and $v^\ast$. The idea is based on the Aslamazov and Larkin (AL) prediction\,\cite{Asl75etp} of kinks in the $I$-$V$ curves of wide and short superconducting constrictions when \emph{the number of vortices crossing the constriction is increased by one}. We realize such conditions in a-few-micrometer-wide MoSi thin strips in which slits, milled by a focused ion beam, locally suppress the edge barrier and allow for a controllable vortex entry. Possessing a high structural uniformity and weak intrinsic pinning, MoSi is a typical SMSPD material\,\cite{Kor14sst,Cal18apl,Kor20pra,Cha20apl} which also features high vortex velocities\,\cite{Bud22pra,Bez22prb,Hab22tsf}. While SMSPDs are usually made from $3$--$13$\,nm-thick films\,\cite{Zha16prb,Chi20apl,Lij16ope,Cap17apl,Gou19ope,Pol20nim,Kor20pra,Cha20apl,Cal18apl,Ust20jpd}, here we use $15$\,nm-thick films to demonstrate the merit of this technique for resolving the open questions for MoSi strips of the same thickness, related to the failure of the $v^\ast(B)$ fits at $B\rightarrow0$\,\cite{Bud22pra} and the unknown number of vortices in the strips at $B=0$\,\cite{Bez22prb}. By measuring the $I$-$V$ curves at zero magnetic field, we observe voltage kinks (minima in the differential resistance) of the strips with increase of the transport current. By associating the kink number $n$ with the number of vortices $n_\mathrm{v}$, we conclude that $n_\mathrm{v}$ varies from one to five in our experiments. From the last point before the FFI jump in the $I$-$V$ curves, at $T = 5$\,K, we deduce the maximal vortex velocity $v^\ast\simeq 12$\,km/s yielding the quasiparticle energy relaxation time $\tau_\epsilon\simeq 30$\,ps. Our experimental observations are augmented with results of the time-dependent Ginzburg-Landau (TDGL) equation modeling which reveals a vortex-chain--vortex-jet--vortex-river transition with increase of the number of vortices and their velocity. An essential result of the TDGL simulations is that kinks in the $I$-$V$ curve are revealed not only for the 1D vortex-chain state, as considered by AL\,\cite{Asl75etp}, but also for the 2D vortex-jet regime\,\cite{Bez22prb}. In all, our findings provide an approach for the deduction of maximal vortex velocities in superconductors and open up a pathway to fluxonic devices in which the number of vortices is controlled by the transport current.
\begin{figure}[b!]
    \centering
    \includegraphics[width=0.98\linewidth]{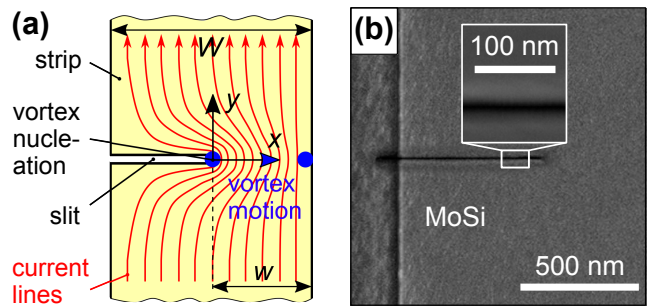}
    \caption{(a) Schematic of a superconducting strip with a slit causing a nonuniform current-density distribution.
    Vortices enter the strip at the slit apex, cross it under the action of the transport current, and exit via the opposite strip side.
    $W$: strip width; $w$: isthmus width. (b) SEM image of a part of the slitted strip MoSi214.
    }
    \label{f1}
\end{figure}

\section{Experiment}

Experiments were carried out on slitted microstrips fabricated from the same $15$\,nm-thick MoSi film. The experimental geometry is shown in Fig.\,\ref{f1}(a). Due to the current-crowding effect\,\cite{Fri01prb,Cle11prb,Ada13apl}, a slit causes a higher current density at its apex and thus determines the place for the entry of vortices into the strip at $I>I_\mathrm{c}$. Here, $I_\mathrm{c}$ is the critical current corresponding to the first vortex entry. Under the action of the transport current, vortices cross the strip and exit it via the opposite side. We note that the significantly stronger edge barrier at the opposite (straight) edge prevents the penetration of antivortices into the strip. This circumstance makes the considered problem distinct from the Glazman model\,\cite{Gla86ltp} dealing with two defects at the opposite strip edges.

\begin{figure*}[t!]
    \centering
    \includegraphics[width=0.98\linewidth]{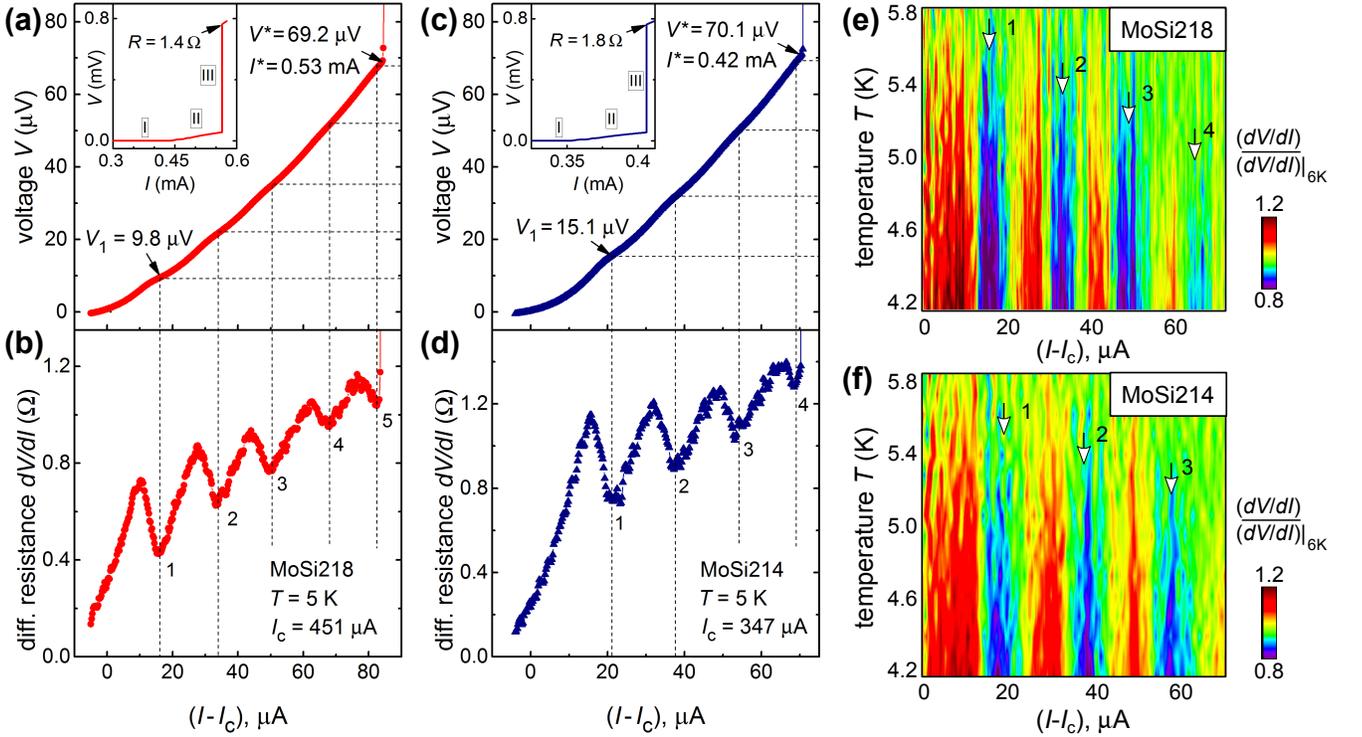}
    \caption{Voltage kinks in the $I$-$V$ curves of the slitted strips MoSi218 (a) and MoSi214 (c) at $T = 5$\,K.
    Inset: The same $I$-$V$ curves in a broader range of currents.
    (b,d) Current dependence of the differential resistance for the same samples.
    (e,f) Smearing of the minima (indicated by the arrows) in the normalized differential resistance $(dV/dI)/(dV/dI)|_\mathrm{6\,K}$ with increase of temperature.
    }
    \label{f2}
\end{figure*}

The MoSi film was deposited by DC magnetron co-sputtering of molybdenum and silicon targets in an Ar atmosphere. The deposition was done onto a Si/SiO$_2$ substrate at room temperature. The deposition rates were adjusted to result in a Mo$_{70}$Si$_{30}$ composition. Further details on the film fabrication and characterization can be found elsewhere\,\cite{Bud22pra}. After the photolithography step, constrictions of $2\,\mu$m width were formed by a focused Ga ion beam (FIB) for electrical resistance measurements. Then, slits with dimensions of $25$\,nm$\times 0.2\,\mu$m and $25$\,nm$\times 0.6\,\mu$m (width$\times$length) were milled by FIB in the middle of one side of the strips labeled MoSi218 and MoSi214, respectively. The slit width was chosen as a minimal width achievable with the FIB milling, to ensure a sharp bending of the current-density lines close to the slit apex. The isthmus widths $w = 1.8\,\mu$m and $1.4\,\mu$m were chosen to result in an at least by $10\%$ smaller critical current $I_\mathrm{c}$ in the slit area with respect to $I_\mathrm{c}$ in the regions away from it and, thus, to exclude the penetration of vortices from other possible defects at the strip edges.

A scanning electron microscopy (SEM) image of a part of the slitted strip MoSi214 is shown in Fig.\,\ref{f1}(b). In the absence of a defect, a current flowing in the $y$ direction would produce a vortex motion in the $x$ direction. The configurations of vortices in the slitted strips and the associated features in their $I$-$V$ curves are the subjects of our studies.

MoSi is an amorphous superconductor in the dirty limit\,\cite{Per05prb,Dob12tsf}. It is characterized by a resistivity $\rho_{8\mathrm{K}}\approx 150\,\mu\Omega$cm, critical temperature $T_\mathrm{c} = 6.43$\,K, upper critical field $B_\mathrm{c2}(0)\approx 10.2$\,T, derivative $dB_{\mathrm{c}2}/dT =-2.23$\,T/K near $T_\mathrm{c}$, electron diffusion coefficient $D\approx0.5$\,cm$^2$/s, coherence length $\xi(0) = \sqrt{\hbar D /1.76k_\mathrm{B}T_\mathrm{c}} = 6$\,nm (corresponding to $\xi_\mathrm{c}=\sqrt{1.76}\xi(0)=7.8$\,nm in the TDGL simulations), penetration depth $\lambda(0) = 1.05\cdot10^{-3} \sqrt{\rho_\mathrm{8K} /T_\mathrm{c}} \approx 500\,$nm\,\cite{Kes83prb}, and $\lambda_\mathrm{eff}(0) = \lambda^2(0)/d \approx 16\,\mu$m. The inequality $\xi \ll w$ means that the strips are \emph{wide} while $w < \lambda_\mathrm{eff}$ justifies the applicability of the AL theory\,\cite{Asl75etp}. In this theory, effects of the current-induced magnetic-field variation on the superconducting properties are disregarded.

Electrical transport measurements were done in the temperature range $T = 4.2$-$5.8$\,K (0.65-0.9$T_\mathrm{c}$) in a He bath cryostat in the standard four-probe geometry. The $I$-$V$ curves were acquired in the current-driven regime at zero magnetic field. The differential resistance data were obtained by differentiating the $I$-$V$ curves with a running average over nine data points.

\section{Results}
\subsection{Voltage kinks in the $I$-$V$ curves}

Figure\,\ref{f2}(a) presents the $I$-$V$ curve of the strip MoSi218 at $T = 5$\,K ($0.78T_\mathrm{c}$). The $I$-$V$ curve has a zero-voltage section (I) at low currents, an extended quasi-linear ohmic branch (II) in the flux-flow regime, and an abrupt FFI jump (III) to a higher resistive state (see the inset in Fig.\,\ref{f2}(a)). A detailed inspection of the $I$-$V$ curve in regime (II) reveals \emph{voltage kinks} which represent our main experimental observation. The first-kink voltage $V_1 = 9.8\,\mu$V. Up to five voltage kinks can already be recognized in the $I$-$V$ curve, but even better so in the differential resistance curve shown in Fig.\,\ref{f2}(b). From the rapid onset of the resistance at low voltages, we deduce the critical current $I_\mathrm{c} = 451\,\mu$A by using a $500$\,nV voltage criterion. The last voltage point before the FFI jump to the highly-resistive state, $V^\ast = 69.2\,\mu$V, corresponds to an instability current of $I^\ast \approx 0.53$\,mA$\simeq 1.2I_\mathrm{c}$.

The $I$-$V$ curve of the strip MoSi214 looks qualitatively similar, see panels (c) and (d) in Fig.\,\ref{f2}. It exhibits four voltage kinks before the FFI onset at the instability voltage $V^\ast = 70.1\,\mu$V at the instability current $I^\ast \approx 0.42$\,mA$\simeq 1.2I_\mathrm{c}$, with $I_\mathrm{c} = 347\,\mu$A and the first-kink voltage $V_1 = 15.1\,\mu$V. The smaller number of voltage kinks for the strip MoSi214 is attributed to the narrow isthmus in this sample.

\begin{figure*}[t!]
    \centering
    \includegraphics[width=0.98\linewidth]{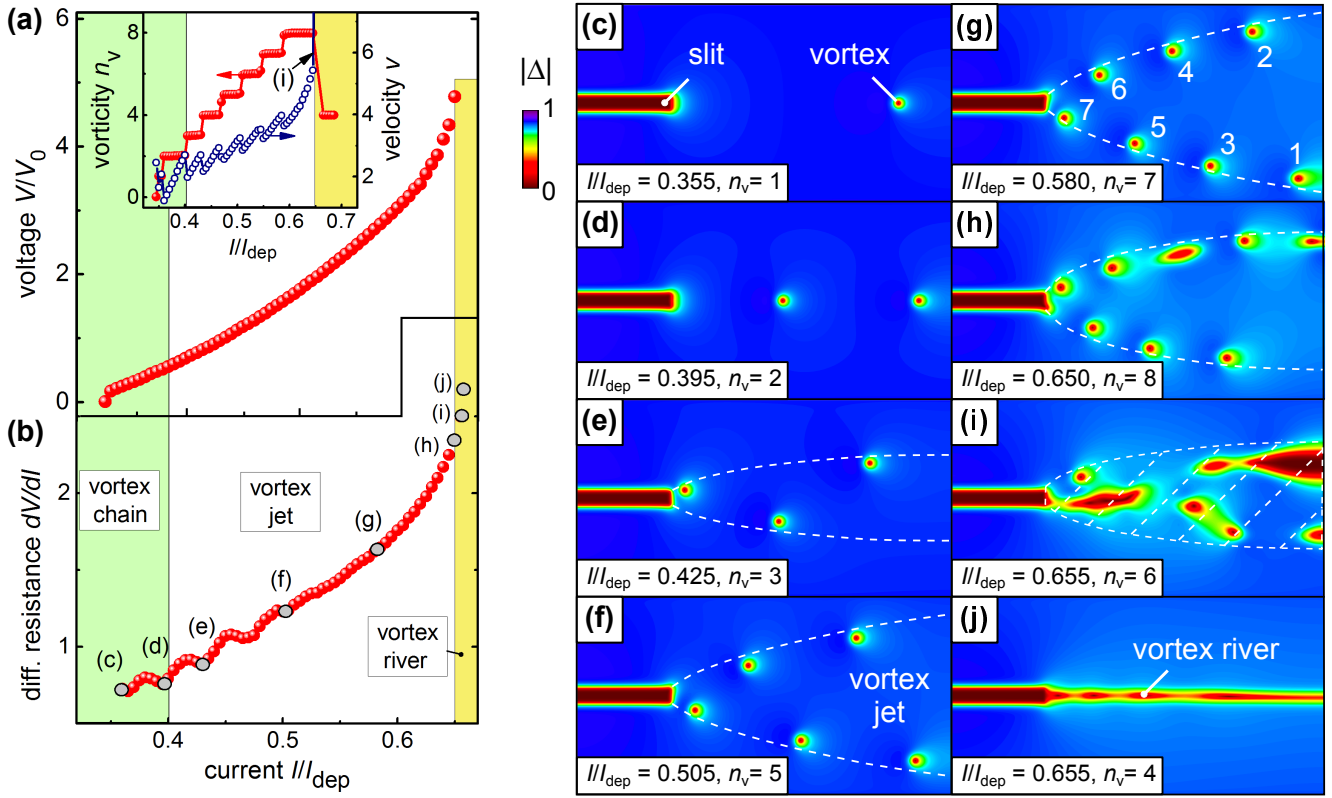}
    \caption{TDGL modeling results.
    (a) $I$-$V$ curve for a superconducting strip with a slit at $T/T_\mathrm{c} = 0.8$. Inset: Evolution of the number of vortices (vorticity) and vortex velocity with increase of the transport current.
    Voltage is in units of $V_0 = 10k_\mathrm{B}T_\mathrm{c}/(2e)$, velocity in units of $10V_0/n_\mathrm{v}$, and current in units of the Ginzburg-Landau depairing current $I_\mathrm{dep}$.
    (b) Current dependence of the differential resistivity.
    (c) Exemplary snapshots (at a given instant of time) of the superconducting order parameter distribution at the $I$-$V$ points indicated in panel (a).
    Panel (i) corresponds to a transient regime between (h) and (j). In the case of strong heating effects, the hatched area in panel (i) is turned into a normal domain which then grows in size until the entire sample transits to the normal state.
    }
    \label{f3}
\end{figure*}

For both samples, the kinks become more pronounced at lower temperatures. As the temperature increases, the kinks are getting smeared and eventually vanish at about $6$\,K. If one takes the current dependence of the differential resistance at 6\,K (without kinks) as a curve for normalization, $(dV/dI)/(dV/dI)|_\mathrm{6\,K}$, the evolution of the kinks upon temperature variation can be seen in panels (e) and (f) in Fig.\,\ref{f2}. Namely, as the temperature increases, the higher-order kinks disappear first and only the first-order kink remains visible at $5.8$\,K.

\subsection{Numerical modeling}
The spatiotemporal evolution of the superconducting order parameter in the strip was modeled on the basis of the TDGL equation\,\cite{Dob20nac,Bud22pra}. The considered equations and the major parameters are detailed in the Appendix. The numerical modeling results are presented in Fig.\,\ref{f3}.

The current dependence of the differential resistance unveils kinks in the $I$-$V$ curve, see Fig.\,\ref{f3}(a,b), just as observed in the experiment in Fig.\,\ref{f2}(a,c). The evolution of the number of vortices in the sample, $n_\mathrm{v}$ (see the inset in Fig.\,\ref{f3}(a)), with increase of the transport current can be understood with the aid of contour plots of the absolute value of the order parameter $|\Delta|$ in Fig.\,\ref{f3}(c)-(h). Namely, the kinks in the flux-flow regime in the $I$-$V$ curve occur when the maximal number of vortices simultaneously crossing the strip is increased by one.

The evolution of the vortex configurations with increase of the transport current can be summarized as follows. At low currents, the vortex trajectories fall onto the same line, forming a vortex chain. When there are more than two vortices in the strip, the vortex chain evolves to a vortex jet. The vortex jet emerges because of the repulsive intervortex interaction\,\cite{Bez22prb}. With increasing vortex velocity, the vortex jet becomes unstable and evolves to a vortex river, see Fig.\,\ref{f3}(i,\,j), that is formed by fast moving Josephson-like vortices. At this transition, the number of vortices $n_\mathrm{v}$ in the strip is reduced by a factor of two while the voltage increases by a factor of three (not shown). These changes in $n_\mathrm{v}$ and $V$ imply a six-fold increase of the vortex velocity. However, heating effects may mask this transition since a normally-conducting domain may appear in the sample (in the hatched region in Fig.\,\ref{f3}(i)) instead of a vortex river. The origin of the voltage kinks and the vortex configurations in the TDGL and AL models will be discussed next.

\section{Discussion}

\subsection{Vortex configurations in the AL and TDGL models}
The central result of our studies is the experimental observation of kinks in the $I$-$V$ curves of slitted superconducting microstrips (Fig.\,\ref{f2}) and their corroboration by the TDGL equation modeling (Fig.\,\ref{f3}). At the same time, our experiments go essentially beyond the AL model due to the 2D geometry. In the AL theory, kinks in the $I$-$V$ curve occur when the number of fluxons in the constriction is increased by one. While the AL theory deals with a 1D chain of vortices along a straight line, our TDGL simulations suggest that this regime is only possible for slits narrower than $\lesssim2\xi_\mathrm{c}$. However, the fabrication of such slits is challenging for superconductors with $\xi(0)\backsimeq5$\,nm.

The slits in our samples are wide ($\simeq3\xi_\mathrm{c}$) which means that the nucleation of every vortex may occur at a slightly different point of the slit apex. As long as there are no more than two vortices in the strip, the vortex-chain arrangement is maintained, see Fig.\,\ref{f3}(d). However, as soon as a third vortex enters into the strip, fluctuations and inhomogeneities, which cannot be avoided even in high-quality samples, facilitate the intervortex repulsion to deflect the trajectories of vortices away from the line $y=0$, thus leading to the formation of a vortex jet\,\cite{Bez22prb}. With an increase of the number of vortices, the opening angle of the vortex jet increases, see Fig.\,\ref{f3}(e)-(g).

In a jet, the vortex trajectories are still very regular. If one assigns a number to each vortex, ``odd'' and ``even'' vortices fall on two separate but well defined lines,  see e.g. Fig.\,\ref{f3}(g). With a further increase of the transport current, the vortex velocity component in the transport current direction ($v_y$) becomes much smaller than the velocity component $v_x$ across the strip so that the opening angle of the vortex jet is decreasing (compare panels (g) and (h) in Fig.\,\ref{f3}). Accordingly, in not very wide strips ($\xi\ll w\ll\lambda$)\,\cite{Bez22prb}, the points where vortices exit the strip at the opposite edge, are displaced toward $y=0$ so that the two vortex rays approach each other. The confluence of the rays gives rise to a vortex river in which the superconducting order parameter is suppressed along it.

A distinct feature of the wider strips ($\xi\lll w\lesssim\lambda$) in our work is that vortex rivers may emerge \emph{in the rays before} they approach each other. Indeed, in the strong-current regime, the TDGL model reveals \emph{halving} of the number of vortices upon a transition to the vortex-river state, see the inset in Fig.\,\ref{f3}(a) where $n_\mathrm{v}$ drops from 8 to 4. This abrupt decrease in the vorticity is indicative of a dynamic topological transition\,\cite{Fom22nsr} accompanied by a drastic increase of the voltage (outside the voltage range shown in Fig.\,\ref{f3}(a)), which is in line with Josephson vortex velocities\,\cite{Emb17nac}.

Alternatively, the transition to the normal state could be mediated by the growth of a normal domain\,\cite{Gur87rmp}. To exclude this mechanism, we have checked that for a sheet resistance of $R_\square\backsimeq80\,\Omega$, a $2\xi \sim 10$\,nm-wide domain would result in at least an order of magnitude larger voltage than measured experimentally and expected for a vortex river. In analogy to phase-slip lines exhibiting Josephson properties in Ref.\,\cite{Siv03prl}, the appearance of Shapiro steps in the microwave-irradiated $I$-$V$ curves of slitted strips in conjunction with a detection of the Josephson generation could be suggested for an independent experimental proof of the vortex-river state in follow-up works.

\subsection{Voltage kinks in the AL versus TDGL models}

\begin{figure}[t!]
    \centering
    \includegraphics[width=0.72\linewidth]{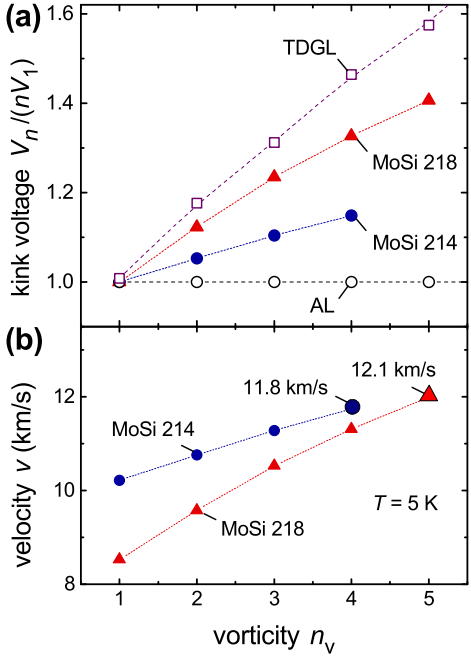}
    \caption{(a) Kink voltage versus number of vortices in the slitted MoSi strips.
    Open symbols: TDGL and AL model calculations.
    Solid symbols: experimental data.
    (b) Vortex velocity versus number of vortices and the $v^\ast$ values (larger symbols) deduced by using Eqs.\,\eqref{eVel} and\,\eqref{eVstar}.
    In both panels, dashed lines are guides for the eye.
    }
    \label{f4}
\end{figure}

The simulations reveal that the kinks in the $I$-$V$ curves are maintained even when a 2D vortex jet is formed. The kinks appear as a consequence of the vortex-vortex repulsion, and the vortices present in the strip hamper the entrance of a new vortex qualitatively in the same manner as in the 1D vortex chains considered by AL\,\cite{Asl75etp}. The quantitative differences between the AL and TDGL models are the following. First, when approaching the kink, the second derivative $d^2V/dI^2$ has a different sign in both models. Second, while the voltage difference between the kinks follows $\Delta V = V_{n+1}-V_n = const$ in the AL model, $\Delta V$ depends on $n$ in the TDGL model, see Fig.\,\ref{f4}(a). The equidistant voltage spectrum in the AL theory is based on the assumption $I_\mathrm{c} \approx I \approx I^\ast$, which implies a steep $I$-$V$ branch and that the voltage increases solely due to the increase of $n_\mathrm{v}$ while the vortex velocity $v$ is assumed to be constant. By contrast, the TDGL model is free of such a restriction regarding the transport current. The kink voltage increase in the TDGL theory is caused by the combined effects of the increase of both the vortex number $n_\mathrm{v}$ and the vortex velocity $v$, see the inset in Fig.\,\ref{f3}(a).

Figure\,\ref{f4}(a) suggests that the kink voltage for the strip MoSi 214 better follows the AL model while for the strip MoSi218 it is better described by the TDGL model. For both our samples $I_\mathrm{c}\gtrsim 0.82I^\ast$, i.e. the comparison with the AL theory is justified. We would like to note that in a different study\,\cite{Bev23rrl} at $T=4.2$\,K, we observed \emph{equidistant} voltage kinks for $4\,\mu$m-wide MoSi strips with $25$\,nm$\times 2\,\mu$m slits. However, those wider constrictions are characterized by $I_\mathrm{c}\approx 0.5I^\ast$, which is rather far from $I_\mathrm{c}\approx I^\ast$ assumed in the AL model. At the moment, we have no explanation for that observation, leaving it for a discussion elsewhere.

Finally, referring to the TDGL simulations at $T=0.8 T_\mathrm{c}$, we note that the discontinuities in the modeled $dV(I)/dI$ curve are smoother than in the experiment. We have checked that these differences are not connected with the vortex jet in our simulations since they also persist for a long and narrow ($\lesssim2\xi_\mathrm{c}$) slit in an identical strip in which a vortex jet appears when the number of vortices in the chain exceeds five. The smoothness of the minima in the $dV(I)/dI$ curve is attributed to the choice of the parameter $\alpha=1$ in the TDGL equation\,\eqref{eTDGL}, whose role is analyzed by Vodolazov and Ustavschikov in their forthcoming work\,\cite{Vod22prv} where the $I$-$V$ curves for slitted MoN thin strips exhibit similar voltage kinks.

\subsection{Vortex counting and velocimetry}
A remarkable feature of the $I$-$V$ curves in Fig.\,\ref{f2}(a) is that the kinks are seen up to the FFI onset. Hence, the kink number $n$ and kink voltage $V_n$ can be used both for counting the number of vortices $n_\mathrm{v}$ and for velocimetry. Both quantities can be deduced for all currents in the flux-flow regime. For example, at $V = V_1$, a transition occurs from less than one vortex to less than two vortices crossing the strip. This means that at any time at $V=V_1$, there is \emph{exactly one vortex}  ($n_\mathrm{v} = 1$) and the average vortex velocity is $v = wV_1/\Phi_0$. Extending this relation to any $V = V_n$, one obtains the generalized expression
\begin{equation}
\label{eVel}
    v = \displaystyle\frac{w V_n}{n_\mathrm{v}\Phi_0}.
\end{equation}

We note that at intermediate voltages $V\neq V_1,V_2,..$, the situation is more complicated because there could be $n$ or $(n+1)$ vortices in the strip.

The instability velocity $v^\ast$ can be deduced from the instability voltage $V^\ast\approx V_n$,
\begin{equation}
\label{eVstar}
    v^\ast = \displaystyle\frac{w V^\ast}{n_\mathrm{v}\Phi_0}.
\end{equation}

In this way, knowing the number of vortices $n_\mathrm{v}$ at each voltage kink, we can deduce the vortex velocity $v$ at $V=V_1,V_2,..$ and make estimates for $v=v^\ast$ at $V=V^\ast$ by using Eqs.\,\eqref{eVel} and\,\eqref{eVstar}. The results of these calculations at $T=5$\,K are presented in Fig.\,\ref{f4}(b). With increase of the current, the vortex velocity increases to $v^\ast$. The deduced values $v^\ast_\mathrm{MoSi218}\approx 12.1$\,km/s and $v^\ast_\mathrm{MoSi214}\approx 11.8$\,km/s are rather close and agree well with the $v^\ast$ values reported for MoSi\,\cite{Bud22pra} from previous FFI studies on straight superconducting strips in the presence of small perpendicular magnetic fields. If we relate the relaxation time $\tau_\varepsilon$ with the healing of the order-parameter wakes behind the moving vortices, $\tau_\varepsilon = w/(v^\ast n_\mathrm{v})$,  we obtain $\tau_\varepsilon^\mathrm{MoSi218}\approx 29$\,ps and $\tau_\varepsilon^\mathrm{MoSi214}\approx 30$\,ps, which also agree well with the previous estimates\,\cite{Bud22pra,Bez22prb}.

The merit of the introduced approach can be illustrated as follows. In principle, one could attempt to estimate the vortex velocity from the self-field $B_\mathrm{sf}$ of the current flowing through the constriction. However, due to the competing effects of the edge barrier and the nonuniform current-density distribution, $B_\mathrm{sf}$ will not necessarily correspond to the actual number of vortices in the strip and, hence, will yield an incorrect vortex velocity. For instance, if one took $I^\ast\approx 0.5$\,mA and $V^\ast\approx 70\,\mu$V at the instability point and modeled the isthmus as a conductor of width $w$ and thickness $d$, the self-field would be $B_\mathrm{sf} = 0.5\,\mu_0 I w^{-1}\ln(2w/d)$, yielding $B_\mathrm{sf} \backsimeq1$\,mT. Assuming that the motion of vortices occurs in a $L\backsimeq50$\,nm-wide isthmus ``channel'', Eq.\,\eqref{e1} yields an unrealistically high $v^\ast = V/(BL) \thickapprox 1400$\,km/s. This overestimation stems from neglecting the strong barrier suppression and the current-crowding in the slit apex area. An accurate calculation of the self-field for geometry-dependent nonuniform current distributions is a complex problem\,\cite{Cle11prb}, which is further complicated by the unknown width of the vortex ``channel''.

The presented approach should be applicable to superconductors with weak vortex pinning, such as NbGe\,\cite{Bab04prb}, MoGe\,\cite{Kun12prb}, WSi\,\cite{Zha16prb}, NbC\,\cite{Dob20nac}, MoN\,\cite{Vod22prv}, and epitaxial Nb films\,\cite{Dob18apl}. Qualitatively, intrinsic vortex pinning is expected to lead to the smearing of the voltage kinks\,\cite{Vod22prv} and to promote the FFI onset at a smaller mean vortex velocity\,\cite{Sil12njp,Dob17sst}.

Finally, we would like to comment on the kinks observed previously in the $I$-$V$ curves of Nb weak links\,\cite{Boo77tom} and YBCO constrictions\,\cite{Niv93prl,Niv94pcs}, which were interpreted within the framework of the AL model. By using Eq.\,\eqref{eVel} one obtains $v^{\mathrm{Nb}}_1=25$\,km/s\,\cite{Boo77tom} and $v^{\mathrm{YBCO}}_1= 35$\,km/s\,\cite{Niv93prl,Niv94pcs} which are by more than an order of magnitude larger than the $v^\ast$ values for these materials\,\cite{Per05prb,Leo11prb,Bez19prb,Doe94prl,Xia96prb,Wor12prb}. We attribute this discrepancy to a different origin of the kinks. For instance, it was argued that the kinks for the YBCO constrictions could be caused by a rearrangement of the moving vortex ensemble\,\cite{Ped95tas}. A detailed inspection of the initial part of the $I$-$V$ curve for those YBCO constrictions (Fig.\,5 in Ref.\,\cite{Ped95tas}) reveals a kink at $V_1 \sim 60\,\mu$V which corresponds to $v_1 \sim 3$\,km/s being in-line with the $v^\ast$ values deduced for YBCO from FFI studies\,\cite{Doe94prl,Xia96prb}. In this way, acquisition of the initial part of the $I$-$V$ curves with a small current step and their careful inspection are essential for the data interpretation.

\subsection{Implications for superconducting devices}
Superconducting microstrips appear as interesting platforms for information processing and sensing. They can find applications in fluxonics, where one can conceive vortex gates, couplers and dividers with several inputs and outputs, as inspired by the vortex configurations in Fig.\,\ref{f3}(g). The placement of Hall voltage leads on both sides of the slit should allow for zero or non-zero transverse voltage outputs depending on whether the vortices cross or do not cross the line between the Hall voltage leads\,\cite{Bez22prb}. The palette of the vortex patterns can be enriched by the fabrication of defects at the opposite strip edge, allowing for steering of vortex-antivortex pairs and their annihilation at a given sample point.

Fast moving fluxons can be used for the Cherenkov-like generation of spin waves\,\cite{Dob19nph,Dob21arx} in superconductor/ferromagnet hybrid systems with prospects for magnon based information processing\,\cite{Chu22tom}. Due to the possible 2D (vortex jet) and 1D (vortex river) arrangements in the ensemble of fast-moving vortices, excitation of two spin-wave beams or a single spin-wave beam could be anticipated.

The high sensitivity of flux steering to local perturbations may also be used to probe the magnetic state of objects placed at the superconductor edge\,\cite{Dob10sst}, which may enable a characteristic entry of fluxons. In this regard, of particular interest are complex-shaped magnetic 3D nanoarchitectures\,\cite{Dob21apl,Mak22adm}, whose characterization by local-field techniques is challenging.

Compared to other superconducting quantum sensors, such as transition edge sensors\,\cite{Irw95apl} and superconducting tunneling junctions\,\cite{Woo69apl}, microwire detectors can be operated at higher temperatures and their response time is orders of magnitude faster than that of kinetic inductance detectors\,\cite{Day03nat}. While superconducting nanowire detectors\,\cite{Sem01pcs} have already found wide application as highly sensitive photon detectors for quantum communication, information processing, infrared astronomy, LIDAR and molecular fluorescence correlation microscopy, our study shows that the dynamics of a single flux quantum can be detected and counted in geometries that are more than two orders of magnitude wider than $\xi$. Relaxation times around 100\,ps would clearly comply with the requirements of fluorescence correlation microscopy, atomic or particle detection, molecular beam science, as well as mass spectrometry\,\cite{Fei19nph}. Microwires are easily integrated to the fabrication of future large-area detectors\,\cite{Ste21apl}.

\section{Conclusion}
Summing up, we have introduced an approach for the quantitative determination of the number of vortices and their velocity in superconducting strips from the $I$-$V$ curves measured at zero magnetic field. The approach is based on the Aslamazov-Larkin prediction of kinks in the $I$-$V$ curves of wide and short constrictions when the number of fluxons crossing the constriction is increased by one. We have fabricated MoSi microstrips with slits as vortex nucleation centers and we have observed a crossover from single- to multi-fluxon dynamics with increase of the transport current. Associating the number of kinks $n$ with the number of vortices $n_\mathrm{v}$ one can calculate the vortex velocity. If the instability voltage reaches $V^\ast\approx V_n$, Eq.\,\eqref{eVstar} can be used to derive the maximal vortex velocity $v^\ast$. If $V^\ast$ notably differs from $V_n$, $v^\ast$ can be estimated by using Eq.\,\eqref{eVstar} with $n_\mathrm{v} = n$ and $n V_1=V \backsimeq V^\ast$. We find that our experimental values at $T = 5$\,K, $v^\ast\simeq12\,$km/s and $\tau_\varepsilon\simeq 30$\,ps agree well with those from prior flux-flow instability studies in the presence of a small perpendicular magnetic field. The TDGL simulations have additionally unveiled a transition from a vortex chain over a vortex jets to a vortex river with increase of the number of vortices. In all, our findings are essential for the development of quantum technologies based on 1D and 2D few-fluxon devices.

\begin{acknowledgments}
The authors are very grateful to Denis Yu. Vodolazov for the TDGL modeling results and numerous fruitful discussions.
V.M.B. acknowledges the European Cooperation in Science and Technology (E-COST) for support via Grants E-COST-STSM-CA19108-48969, E-COST-GRANT-CA16218-5759aa9b, and E-COST-GRANT-CA16218-46e403c7.
M.Yu.M. acknowledges the Wolfgang Pauli Institute (WPI) Vienna for the scholarship within the framework of the Pauli Ukraine Project,
the scholarship from the Krzysztof Skubiszewski Foundation, and the IEEE Magnetics Society for support via the STCU Project No. 9918.
B.B. and S.L.C. acknowledge financial support by the Vienna Doctoral School in Physics (VDSP).
S.O.S. acknowledges the OeAD (Austria's Agency for Education and Internationalisation) for support through the Ernst Mach Grant, EM UKR.
CzechNanoLab project LM2018110 is gratefully acknowledged for financial support of the sample fabrication at CEITEC Nano Research Infrastructure.
This research is funded in whole, or in part, by the Austrian Science Fund (FWF), Grant No.\,I\,4865-N.
Support by E-COST via COST Actions CA19108 (HiSCALE) and CA21144 (SuperQuMap) is gratefully acknowledged.
\end{acknowledgments}

\section*{Appendix}
The spatiotemporal evolution of the superconducting order parameter $\Delta=|\Delta|\exp(i\phi)$ in the strips was modeled numerically on the basis of the TDGL equation
\begin{eqnarray}
\label{eTDGL}
    \alpha \frac{\pi\hbar}{8k_\mathrm{B}T_\mathrm{c}} \left(\frac{\partial }{\partial t}+\frac{2ie\varphi}{\hbar} \right) \Delta= \nonumber
    \\
    \frac{\pi}{8}\xi_\mathrm{c}^2\left( \nabla -i\frac{2e}{\hbar c}A\right)^2\Delta+\left(1-\frac{T}{T_\mathrm{c}}-\frac{|\Delta|^2}{\Delta_\mathrm{GL}^2}\right)\Delta
\end{eqnarray}
coupled with the equation for the electrostatic potential
$\varphi$
\begin{equation}
\label{eEP}
    \mathrm{div} \,\mathbf{j} = \mathrm{div} \,(\mathbf{j}_\mathrm{s}+\mathbf{j}_\mathrm{n})=
    \mathrm{div} \left(\frac{\sigma_\mathrm{n}}{e} \frac{\pi|\Delta|^2q_\mathrm{s}}{4k_\mathrm{B}T_\mathrm{c}}-\sigma_\mathrm{n}\nabla\varphi \right) = 0,
\end{equation}
which follows from the continuity of the current lines.

In Eqs.\,\eqref{eTDGL} and \eqref{eEP}, $\phi$ is the phase of the superconducting order parameter, the phenomenological coefficient $\alpha$ controls the relaxation time of the superconducting order parameter, $\tau_{|\Delta|}\sim \alpha$, $T_\mathrm{c}$ is the superconducting transition temperature, $\xi_\mathrm{c}=\sqrt{\hbar D/k_\mathrm{B}T_\mathrm{c}}$, $\sigma_\mathrm{n} = 2 e^2 D N_0$ is the normal-state conductivity, $N_0$ is the electron density of states on the Fermi surface per electron spin, $\Delta_\mathrm{GL} = 3.06 k_\mathrm{B}T_\mathrm{c}$, $A$ is the vector potential, and $q_\mathrm{s} = (\nabla \varphi - 2 e A / \hbar c)$ is proportional to the supervelocity.

Details on the procedure employed for solving Eqs.\,\eqref{eTDGL} and \eqref{eEP} were reported in Refs.\,\cite{Bud22pra,Ust22etp}. In contrast to Refs.\,\cite{Bud22pra,Ust22etp}, the TDGL in the present study was solved in neglect of overheating effects, since these are only essential very close to the FFI point (regime III). This simplification is justified since the central observation of our studies (kinks in the $I$-$V$ curves) is related to the low-dissipative flux-flow regime (II).

The superconducting strip was modeled as a rectangular polygon with dimensions $200\xi_\mathrm{c}\times 130\xi_\mathrm{c}$ ($x\times y$). The slit was modeled as a rectangular region of size $50\xi_\mathrm{c}\times 7\xi_\mathrm{c}$ ($x\times y$) with a local suppression of $T_\mathrm{c}$. It was also checked that the modeling of wide slits ($\gtrsim2\xi_\mathrm{c}$) as geometrical defects (i.e. no material in the slit region) only leads to small quantitative differences in the $I$-$V$ curves but is much less time-efficient because of the more complicated boundary conditions around the slit. The calculations were done for $\alpha =1$.


%

\end{document}